\documentclass[aps,prl,twocolumn,showpacs]{revtex4-1}
\usepackage[utf8]{inputenc}
\usepackage{graphicx}
\usepackage{amsmath}
\usepackage{hyperref}
\usepackage{units}
\usepackage{amsmath, amssymb}
\usepackage{natbib}

\usepackage{color}

\begin{document}

\title{Exploring Symmetry Breaking at the Dicke Quantum Phase Transition}
\author{K.~Baumann}
\author{R.~Mottl}
\author{F.~Brennecke}
\email{brennecke@phys.ethz.ch}
\author{T.~Esslinger}
\affiliation{Institute for Quantum Electronics, ETH Z\"{u}rich,
8093 Z\"{u}rich, Switzerland}

\begin{abstract}
We study symmetry breaking at the Dicke quantum phase transition
by coupling a motional degree of freedom of a Bose-Einstein
condensate to the field of an optical cavity. Using an optical
heterodyne detection scheme we observe symmetry breaking in
real-time and distinguish the two superradiant phases. We explore
the process of symmetry breaking in the presence of a small
symmetry-breaking field, and study its dependence on the rate at
which the critical point is crossed. Coherent switching between
the two ordered phases is demonstrated.

\end{abstract}
 \maketitle

Spontaneous symmetry breaking at a phase transition is a
fundamental concept in physics \cite{huang1987}. At zero
temperature, it is caused by the appearance of two or more
degenerate ground states in the Hamiltonian. As a result of
fluctuations, a macroscopic system evolves into one particular
ground state which does not possess the same symmetry as the
Hamiltonian. Finding a clean testing ground to experimentally
study the process of symmetry breaking is notoriously difficult as
external fluctuations and asymmetries have to be minimized or
controlled. The protected environment of atomic quantum gas
experiments and the increasing control over these systems offer
new prospects to experimentally approach the concept of symmetry
breaking. Recently, rapid quenches across a phase transition were
studied in multi-component Bose-Einstein condensates \cite{Sadler2006,Kronjager2010,Scherer2010} and optical lattices \cite{Bakr2010,Chen2011}. Such a non-adiabatic quench causes a response of the system at correspondingly high
energies. Therefore, a central characteristic of a phase
transition, which is its diverging susceptibility to
perturbations, remains partially hidden.

In this work we study the symmetry breaking process while slowly
varying a control parameter several times across a
zero-temperature phase transition. Compared to quenching, this
allows us to explore the low energy spectrum of the system which
probes its symmetry most sensitively. For very slow crossing
speeds we identify the presence of a residual symmetry breaking
field of varying strength. Larger values of this residual field
can be correlated to the repeated observation of one particularly
ordered state. For increasingly steeper ramps across the phase
transition the influence of the symmetry breaking field almost
vanishes.

We investigate the symmetry breaking in the motional degree of
freedom of a Bose-Einstein condensate coupled to a single mode of
an optical cavity. Our system realizes the Dicke model
\cite{Dicke1954,Baumann2010,Nagy2010} which exhibits a
second-order zero-temperature phase transition
\cite{Hepp1973,Wang1973,Carmichael1973,Emary2003}. The broken
symmetry is associated with the formation of one of two identical
atomic density waves, which are shifted by half an optical
wavelength \cite{Baumann2010,Nagy2010,Domokos2002,Black2003}.
Using an interferometric heterodyne technique, we monitor the
symmetry-breaking process in real time while crossing the
transition point. A similar technique has been used to test
self-organization in a classical ensemble of laser-cooled atoms
\cite{Black2003}, where the symmetric phase is stabilized by
thermal energy rather than kinetic energy \cite{Nagy2008}.

%%%%%%%%%%%%%%%%%%
\begin{figure}[b!]
\includegraphics[width=1\columnwidth]{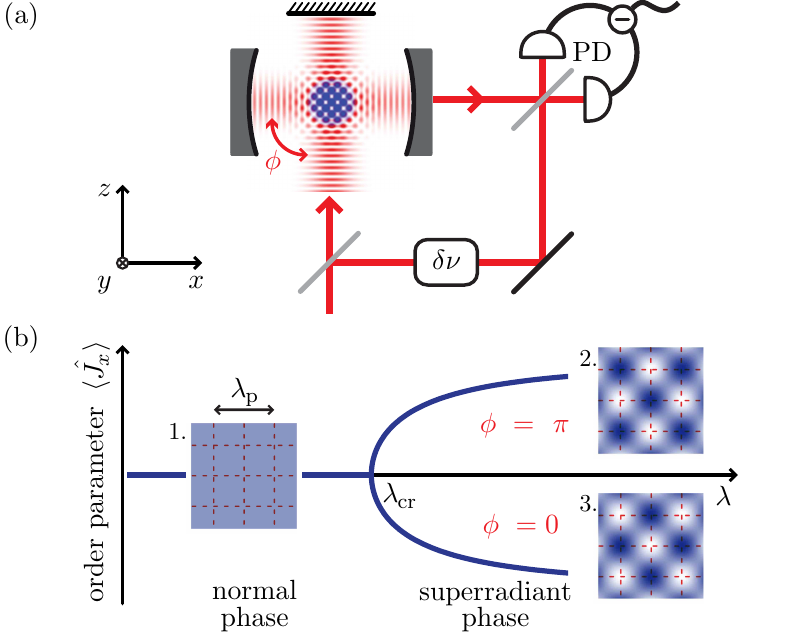}
\caption{(Color online) (a) Experimental setup. A Bose-Einstein
condensate is placed inside an optical cavity and driven by a
far-detuned standing-wave laser field (wavelength
$\lambda_\mathrm{p}$) along the $z$-axis. Phase and amplitude of
the intracavity field are measured with a balanced heterodyne
setup (PD: photodiodes). (b) Steady-state order parameter $\langle
\hat{J}_x \rangle$ as a function of coupling strength $\lambda$,
with corresponding atomic density distributions (1.-3.). The order
parameter vanishes in the normal phase (1.) and bifurcates at the
critical point $\lambda_\mathrm{cr}$, where a discrete
$\lambda_\mathrm{p}/2$-spatial symmetry is broken. The two
emergent superradiant phases (2. and 3.) can be distinguished via
the relative time-phase $\phi$.} \label{fig1}
\end{figure}
%%%%%%%%%%%%%%%%%%

The Dicke model \cite{Dicke1954} considers the interaction between
$N$ two-level atoms and the quantized field of a single-mode
cavity, which is described by the Hamiltonian
\begin{equation}
\hat{H} = \hbar\omega_0 \hat{J}_z + \hbar\omega\hat{a}^\dag\hat{a}
+ \frac{2\hbar \lambda}{\sqrt{N}}(\hat{a}^\dag+\hat{a})\hat{J}_x.
\label{eq:dicke_hamiltonian}
\end{equation}
Here, $\hat{a}$ and $\hat{a}^\dag$ denote the annihilation and
creation operators for the cavity mode at frequency $\omega$, and
$\hat{\mathbf{J}} = (\hat{J}_x,\hat{J}_y,\hat{J}_z)$ describes the
atomic ensemble with transition frequency $\omega_0$ in terms of a
pseudospin of length $N/2$. The cavity light field couples with
coupling strength $\lambda$ to the collective atomic dipole
$\hat{J}_x$. In the thermodynamic limit, the Dicke model exhibits
a zero-temperature phase transition from a normal to a
superradiant phase when the control parameter $\lambda$ exceeds a
critical value given by $\lambda_\mathrm{cr} = \sqrt{\omega
\omega_0}/2$ \cite{Hepp1973,Wang1973,Carmichael1973}.
Simultaneously, the parity symmetry of the Dicke Hamiltonian,
given by the invariance under the transformation $(\hat{a},
\hat{J}_x) \rightarrow (-\hat{a}, -\hat{J}_x)$, is spontaneously
broken \cite{Emary2003}. While parity is conserved in the normal
phase with $\langle \hat{a} \rangle = 0 = \langle \hat{J_x}
\rangle$, two equivalent superradiant phases (denoted by even and
odd) emerge for $\lambda > \lambda_\mathrm{cr}$, which are
characterized by $\langle \hat{J}_x \rangle \lessgtr 0$ and
$\langle \hat{a} \rangle \gtrless 0$, respectively
(Fig.~\ref{fig1}b).

In our experiment \cite{Baumann2010} we couple motional degrees of
freedom of a Bose-Einstein condensate (BEC) with a single cavity
mode using a transverse coupling laser (Fig.~\ref{fig1}a). Within
a two-mode momentum expansion of the matter-wave field, the
Hamiltonian dynamics of this system is described by the Dicke
model (Eq.~\ref{eq:dicke_hamiltonian}) \cite{Baumann2010,Nagy2010,
Dimer2007} where the effective atomic transition frequency is
given by $\omega_0 = 2\omega_\mathrm{r}$ with the recoil frequency
$\omega_\mathrm{r} = \hbar k^2/2m$, the atomic mass $m$ and the
wavelength $\lambda_\mathrm{p} = 2\pi/k$ of the coupling laser.
The frequency and power of this laser controls the effective mode
frequency $\omega$ and the coupling strength $\lambda$,
respectively \cite{Baumann2010}. Above a critical laser power, the
discrete $\lambda_\mathrm{p}/2$-spatial symmetry, defined by the
optical mode structure $u(x,z) = \cos(kx)\cos(kz)$, is
spontaneously broken and the condensate exhibits either of two
density waves (Fig.~\ref{fig1}b). Correspondingly, the atomic
order parameter $\langle \hat{J_x} \rangle$, given by the
population difference between the even ($u(x,z)>0$) and odd
($u(x,z)<0$) sublattice, exhibits a negative or positive
macroscopic value, while the emergent coherent cavity field
oscillates (for $\omega \gg \kappa$) either in ($\phi = 0$) or out
of phase ($\phi = \pi$) with the coupling laser.

\begin{figure}[]
\includegraphics[width=1\columnwidth]{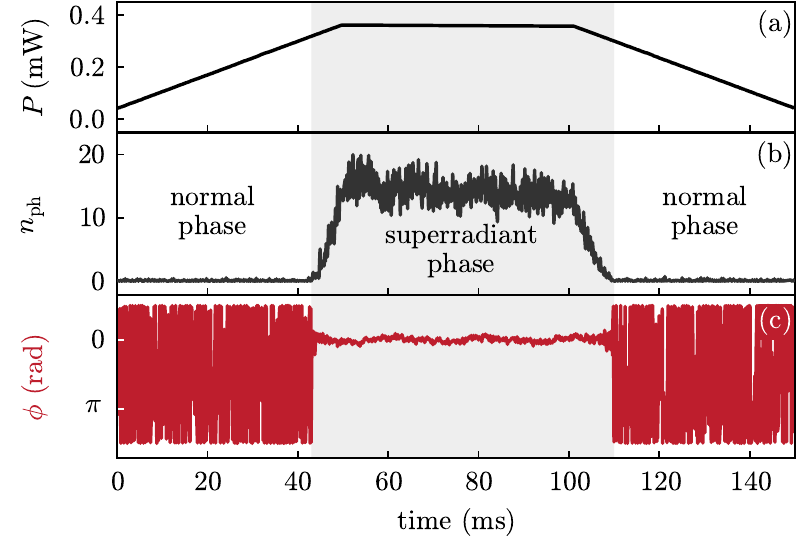}
\caption{(Color online) Observation of symmetry breaking and
steady-state superradiance. Shown are simultaneous traces of (a)
the coupling laser power $P$, (b) the mean intracavity photon
number $n_\mathrm{ph}$, and (c) the relative time-phase $\phi$
between coupling laser and cavity field (both averaged over
$\unit[150]{\mu s}$). The coupling laser frequency is red-detuned
by $\unit[31.3(2)]{MHz}$ from the empty cavity resonance and the
atom number is $2.3(5)\times 10^5$. Residual atom loss causes a
slight decrease of the cavity photon number in the superradiant
phase.} \label{fig2}
\end{figure}

As described previously \cite{Baumann2010}, we prepare BECs of
typically $2 \times 10^5$ $^{87}\mathrm{Rb}$ atoms in a
crossed-beam dipole trap centered inside an ultrahigh-finesse
optical Fabry-Perot cavity, which has a length of $\unit[176]{\mu
m}$. The transverse coupling laser at wavelength
$\lambda_\mathrm{p} = \unit[784.5]{nm}$ is red-detuned by
typically ten cavity linewidths $2 \kappa = 2\pi \times
\unit[2.5]{MHz}$ from a $\mathrm{TEM}_{00}$ cavity mode, realizing
the dispersive regime $\omega \gg \omega_0$ of the Dicke model. We
monitor amplitude and phase of the intracavity field in real-time
using a balanced heterodyne detection scheme (Fig.~\ref{fig1}a).
Due to slow residual drifts of the differential path length of our
heterodyne setup, which translate into drifts of the detected
phase signal of about $\unit[0.1\pi]{/s}$, we cannot relate the
phase signals between consecutive experimental runs separated by
$\unit[60]{s}$.

To observe symmetry breaking, we gradually increase the coupling
laser power across the critical point (Fig.~\ref{fig2}a). The
transition from the normal to the superradiant phase is marked by
a sharp increase of the mean intracavity photon number
(Fig.~\ref{fig2}b). Simultaneously, the time-phase $\phi$ between
the two light fields locks to a constant value, implying that the symmetry of the
system has been broken (Fig.~\ref{fig2}c). The observation of a
constant time-phase above threshold confirms that the system
reaches a steady-state superradiant phase in which the induced
cavity field oscillates at the coupling laser frequency. When
lowering the laser power to zero again, the system recovers its
initial symmetry and a pure BEC is retrieved, as was inferred from
absorption imaging after free ballistic expansion.

\begin{figure}[t]
\includegraphics[width=1\columnwidth]{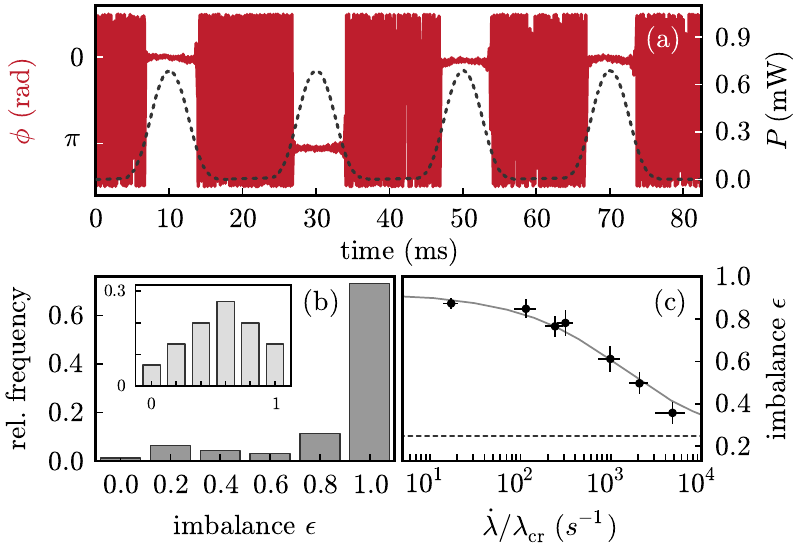}
\caption{(Color online) (a) Cavity time-phase (red, averaged over
$\unit[30]{\mu s}$) for a single run, and corresponding time
sequence of the coupling laser power $P$ (dashed). (b) Probability
distribution of the imbalance $\epsilon$ (see text) for 156 runs,
where the phase transition was crossed at a rate of
$\dot{\lambda}/\lambda_\mathrm{cr} = \unit[18(3)]{s^{-1}}$. The
inset displays the distribution of post-selected data (see text).
(c) Mean imbalance (dots) as a function of the rate
$\dot{\lambda}/\lambda_\mathrm{cr}$ at which the transition was
crossed (extracted from 356 runs in total), and theoretical model
(solid line). The error bars indicate the standard error of the
mean of $\epsilon$ and systematic changes of $\lambda_\mathrm{cr}$
during probing. } \label{fig3}
\end{figure}

To identify the two different superradiant states
(Fig.~\ref{fig1}b), we cross the phase transition multiple times
within one experimental run (Fig.~\ref{fig3}a). Above threshold,
the corresponding phase signal takes always one of two constant
values. From multiple traces of this type we extract a time-phase
difference of $1.00(2)\times \pi$ between the two superradiant
phases, where the statistical error can be attributed to residual
phase drifts of our detection system.

If the system was perfectly symmetric, the two ordered phases
would be realized with equal probabilities, when repeatedly
crossing the phase transition. However, the presence of any
symmetry-breaking field will always drive the system into the same
particularly ordered state when adiabatically crossing the
critical point. We experimentally quantify the even-odd imbalance
by performing 156 experimental runs (similar to Fig.~\ref{fig3}a),
in each of which the system enters the superradiant phase ten
times within $\unit[1]{s}$. A measure for the even-odd imbalance
is given by the parameter $\epsilon = (m_1-m_2)/10$, where
$m_2\leq m_1$ denote the number of occurrences of the two
superradiant configurations in individual traces. In
$\unit[73]{\%}$ of the traces, the system realized ten times the
same time-phase, corresponding to the maximum imbalance of
$\epsilon = 1$ (Fig.~\ref{fig3}b). However, $\unit[12]{\%}$ of the
runs exhibited an imbalance below 0.5, which is not compatible
with a constant even-odd asymmetry.

We attribute our observations to the finite spatial extension of
the atomic cloud. This can result, even for zero coupling
$\lambda$, in a small, but finite population difference between
the even and odd sublattice, determined by the spatial overlap
$\mathcal{O}$ between the atomic column density $n(x,z)$
(normalized to $N$) and the optical mode profile $u(x,z)$. This
asymmetry enters the two-mode description
(Eq.~\ref{eq:dicke_hamiltonian}) via the symmetry-breaking term
$2\hbar \lambda \mathcal{O} (\hat{a}^\dag + \hat{a}) /\sqrt{N}$,
and renormalizes the order parameter $\langle \hat{J}_x \rangle$
by the additive constant $\mathcal{O}$. The resulting coherent
cavity field below threshold drives the system dominantly into
either of the two superradiant phases, depending on the sign of
$\mathcal{O}$. In the experiment, the resulting even-odd imbalance
is likely to change between experimental runs, as the overlap
integral $\mathcal{O}$ depends $\lambda_p$-periodically on the
relative position between the mode structure $u(x,z)$ and the
center of the trapped atomic cloud, with amplitude
$\mathcal{O}_0$. We can exclude a drift of the relative trap
position by more than half a wavelength $\lambda_\mathrm{p}$ on
the timescale given by our probing time of $\unit[1]{s}$, as it
would lead to equal probabilities of the two phases, pretending
spontaneous symmetry breaking.

The openness of the system provides us with direct experimental
access to the symmetry-breaking field proportional to
$\mathcal{O}$. Indeed, we detect a small coherent cavity field
($n_\mathrm{ph} < 0.02$) in the normal phase whose magnitude
varies between experimental runs. In all runs exhibiting an
imbalance of $\epsilon = 1$ (Fig.~\ref{fig3}b), the relative
time-phases of the cavity field detected below and above threshold
are equal. Furthermore, the even-odd imbalance increases
significantly with the light level observed below threshold.
Post-selection of those $\unit[10]{\%}$ of the runs with the
smallest light level yields a much smaller mean imbalance
(Fig.~\ref{fig3}b, inset).

In general, the influence of a symmetry breaking field becomes
negligible, if the mean value of the order parameter, induced by
this field, is smaller than the quantum or thermal fluctuations
present in the system. From a mean-field calculation performed in
the Thomas-Fermi limit for $N = 2\times 10^5$ harmonically trapped
atoms, we estimate a maximum order parameter of $\mathcal{O}_0 =
40$ for zero coupling strength, corresponding to an even-odd
population difference of 40 atoms. This value is much smaller than
the uncertainty $\Delta J_x = \sqrt{N}/2 = 224$, given by vacuum
fluctuations of the excited momentum mode. Therefore, one expects
in the extreme case of a sudden quench of the coupling strength
beyond $\lambda_\mathrm{cr}$, that the apparent symmetry is
spontaneously broken, resulting in nearly equal probabilities of
the two superradiant phases.

In the experiment we determined the even-odd imbalance $\epsilon$
for increasingly larger rates $\dot{\lambda}/\lambda_\mathrm{cr}$
at which the critical point was crossed, i.e.~in an increasingly
non-adiabatic situation (Fig.~\ref{fig3}c). As the transition is
crossed faster, the mean imbalance between the two superradiant
phases decreases significantly and approaches the value $\epsilon
\approx 0.25$ corresponding to the balanced situation
(Fig.~\ref{fig3}c, dashed line). This indicates that the effect of
the symmetry breaking term can be overcome by non-adiabatically
crossing the phase transition.

Our observations (Fig.~\ref{fig3}c) are in quantitative agreement
with a simple model based on the adiabaticity condition known from
the Kibble-Zurek theory \cite{Zurek2005, Dziarmaga2010}. We divide
the evolution of the system during the increase of the transverse
laser power into a quasi-adiabatic regime, where the system
follows the change of the control parameter, and an impulse
regime, where the system is effectively frozen. After crossing the
critical point, fluctuations of the order parameter, which are
present at the instance of freezing, become instable and are
amplified. The coupling strength which separates the two regimes
is determined by Zurek's equation \cite{Zurek2005}
$|\dot{\zeta}/\zeta| = \Delta/\hbar$, with $\zeta =
(\lambda_\mathrm{cr}-\lambda)/\lambda_\mathrm{cr}$ and the energy
gap between ground and first excited state given by $\Delta =
\hbar \omega_0 \sqrt{1-\lambda^2/\lambda_\mathrm{cr}^2}$ for
$\omega \gg \omega_0$ \cite{Emary2003, Dimer2007}.

We deduce the probability with which the system chooses the even
phase, $p_\mathrm{even} = \int_0^\infty p(\Theta) \mathrm{d}
\Theta$, from the probability distribution $p(\Theta)$ at the
instance of freezing, where $\hat{\Theta}$ denotes the shifted
dipole operator $\hat{\Theta} = \hat{J}_x + \mathcal{O}$. In the
thermodynamic limit the distribution $p(\Theta)$ becomes Gaussian
with a mean value $\langle \hat{\Theta} \rangle = \langle
\hat{J}_x\rangle + \mathcal{O}$ and a width determined by the
quantum fluctuations of the order parameter $\Delta J_x$. These
values are determined from the linear quantum Langevin equations
based on the Dicke model \cite{Dimer2007} including the symmetry
breaking term. Besides the decay of the cavity field we also take
into account dissipation of the excited momentum state at a rate
$\gamma = 2\pi\times \unit[0.6]{kHz}$. This value was deduced from
independent measurements of the cavity output field below
threshold \cite{cavity2011}.

From the steady-state solution of the quantum Langevin equations
we find that the mean order parameter $\langle \hat{\Theta}
\rangle$ grows faster in $\lambda$ than its fluctuations. If
$\mathcal{O}>12$ the order parameter exceeds its uncertainty
already below critical coupling. Thermal fluctuations are
neglected in this analysis. For our typical condensate
temperatures of about $\unit[100]{nK}$ quantum fluctuations
dominate as long as $\zeta > 0.005$. We account for shot-to-shot
fluctuations of the overlap $\mathcal{O}$ by suitably averaging
over the position of the harmonic trap. The solid line in
Fig.~\ref{fig3}c shows a least square fit of our model to the data
with the single free parameter $\mathcal{O}_0$. We obtain a value
of $\mathcal{O}_0 = 77$ which is in reasonable agreement with the
theoretically expected value of $\mathcal{O}_0 = 40$. This
verifies the predominance of the considered symmetry breaking
field over other possible noise terms.

\begin{figure}[t!]
\includegraphics[width=1\columnwidth]{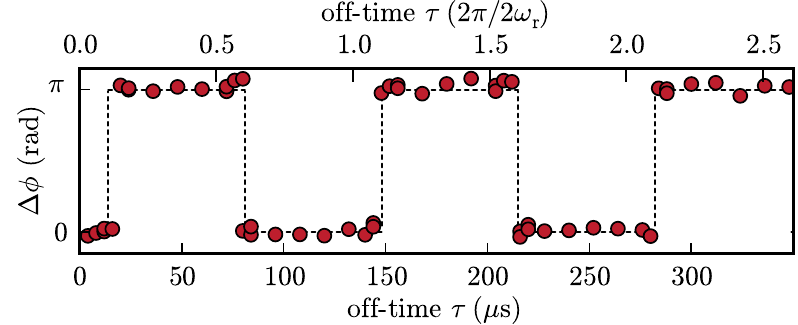}
\caption{(Color online) Coherent switching between the two ordered
phases. After adiabatically preparing the system in one of the two
superradiant phases, the coupling field is turned off for a time
$\tau$. Displayed is the steady-state cavity time-phase $\Delta
\phi$ (averaged over $\unit[0.5]{ms}$) after turning on the
coupling field, referenced to the value recorded before the
turning-off. Each data point corresponds to a single measurement.
The dashed line shows the time evolution as expected from the
two-mode model.} \label{fig4}
\end{figure}

Finally, we experimentally demonstrate coherent switching between
the two ordered states. To this end we suddenly turn off the
coupling laser field after adiabatically preparing the system in
one of the two superradiant phases. The atoms are then allowed to
freely evolve according to their momentum state occupation, giving
rise to standing-wave oscillations of the atomic density
distribution. In the two-mode description this corresponds to
harmonic oscillations of the order parameter $\langle
\hat{J}_x\rangle$ at frequency $2\omega_\mathrm{r}$. We probe this
time evolution by turning on the coupling laser after a variable
off-time $\tau$, thereby deterministically re-trapping the atoms
either in the initial or in the opposite superradiant state. As
expected, we observe regular $\pi$-jumps in the difference $\Delta
\phi$ between the steady-state phase signals measured before and
after the free evolution, with a frequency of $2\omega_\mathrm{r}$
(Fig.~\ref{fig4}, dashed line). The inertia of the atoms traveling
at finite momentum causes the $\pi$-jumps in Fig.~4 to occur
before those times at which the order parameter has evolved by an
odd number of quarter periods.

In conclusion, we have experimentally monitored symmetry breaking
in the Dicke quantum phase transition and identified the interplay
between a residual symmetry breaking field, fluctuations and the
crossing speed.

We acknowledge discussions with G.~Blatter, P.~Domokos, T.~Donner,
R.~Landig, B.~Oztop, L.~Pollet, H.~Ritsch, and H.~Tureci.
Financial funding from NAME-QUAM (European Union), SQMS (ERC) and
QSIT (NCCR) is acknowledged.

\end{document}